# Quantum Inspired QUBO Assisted ALNS for Reliability Driven Hurricane Restoration of Distribution Networks

Hooman Torkaman, Jignesh Solanki, and Anurag Srivastava
Lane Department of Computer Science and Electrical Engineering, West Virginia University, Morgantown, WV, USA

Abstract: Post hurricane distribution system restoration requires rapid repair scheduling subject to feeder topology, field logistics, and electrical feasibility. This paper presents a quantum inspired quadratic unconstrained binary optimization (QUBO) assisted adaptive large neighborhood search (ALNS) framework. At each restoration stage, a local CPU simulated annealing sampler ranks individual repairs and multi job combinations near the energized frontier. A deterministic decoder preserves crew truck logistics, enforces full useful crew utilization, and rejects infeasible batches. Final schedules are validated through OpenDSS replay. The framework is evaluated on the IEEE 123 node test feeder without distributed generation under 80, 90, and 100 m/s wind scenarios. In the 100 m/s stress test, the proposed method reduces mean system average interruption duration index and energy not supplied by 2.24% and restoration makespan by 50.71% relative to classical energized ALNS. Results show that QUBO assistance is most valuable when severe damage creates a larger combinatorial repair space.

## I. Introduction

Extreme weather events can cause sustained outages by damaging overhead distribution assets and isolating large groups of customers. The resilience challenge is not limited to identifying failed components. Utilities must determine which repairs should start first, which activities can proceed in parallel, and how limited field resources should be routed so that service is restored quickly and safely. Weather related outages are especially difficult because failure locations, repair durations, and accessibility constraints are spatially distributed across the feeder [1].

Post event restoration has therefore been studied as a coordinated optimization problem. Repair and resource scheduling models show that restoration quality depends on the order and timing of field activities rather than only the final repaired topology [2]. Adaptive large neighborhood search (ALNS) is attractive for this setting because it can update a candidate schedule through destroy and repair operators while retaining a physically interpretable decoder [3]. However, stage level batch selection remains combinatorial: a scheduler must choose among individual repairs, multi job combinations, and compatible parallel batches while considering upstream downstream dependencies.

Quadratic unconstrained binary optimization (QUBO) provides a compact representation of binary combinatorial decisions [4], [5]. QUBO models are compatible with quantum annealing workflows, but they can also be evaluated classically during development. Simulated annealing remains a useful local sampling mechanism for such binary energy landscapes [6]. Recent power system research has explored quantum annealing compatible restoration formulations, including hybrid quantum classical microgrid formation for critical load restoration [7]. A narrower but practical question remains: can QUBO assisted batch selection improve repair scheduling while leaving trusted utility logistics and OpenDSS validation layers unchanged?

Recent learning based studies provide complementary perspectives on restoration and fault management. An event driven deep reinforcement learning dispatcher has been proposed for post storm crew assignment with repeated replanning and feasibility masking [8], while a Transformer based deep reinforcement learning policy has been developed for post earthquake crew dispatch with power flow re solving and energy not supplied aware evaluation [9]. Residual one dimensional convolutional neural networks have also been applied to distribution network fault classification and line location tasks [10]. These approaches address online dispatch or upstream diagnosis; the present study instead focuses on QUBO assisted stage batch selection inside a logistics preserving ALNS framework.

This paper addresses that question using the IEEE 123 node test feeder [11]. The present implementation is deliberately described as quantum inspired: the QUBO is solved locally on a CPU by simulated annealing and is not submitted to physical quantum hardware. The main contributions are:

1) A QUBO assisted energized ALNS scheduler that evaluates individual repair jobs and multi job combinations within the same stage selection layer.
2) An adaptive candidate pool mechanism that expands beyond the immediate energized frontier when additional productive work is needed, together with a hard full useful crew acceptance rule.
3) A deterministic logistics filter that preserves crew and truck travel, material trip timing, and repair duration logic after QUBO sampling.
4) An OpenDSS validated evaluation on the IEEE 123 node test feeder across three wind severity levels, including reliability, makespan, and runtime comparisons.

## II. Methodology

### A. Spatial Hazard and Damage Model

Bus coordinates are read from the feeder coordinate file and mapped into a local planar coordinate system. Line exposure is evaluated at each line midpoint. A synthetic storm eye travels through the feeder region, and the distance from line midpoint $i$ to the storm eye at time step $t$ is

$$r_{i,t} = \sqrt{\left(x_i - x_t^{\text{eye}}\right)^2 + \left(y_i - y_t^{\text{eye}}\right)^2} \quad (1)$$

To provide a computationally efficient spatial stress profile, the line level wind speed is represented by a radial exponential envelope:

$$v_{i,t} = V_{\max}\exp\left[-\left(\frac{r_{i,t}}{R_w}\right)^2\right] \quad (2)$$

where $V_{max}$ is the scenario maximum wind speed and $R_w$ is the wind field radius. The formulation is a simplified parametric hazard representation intended for restoration scheduling studies; more detailed vortex models can be substituted without changing the scheduling architecture [12].

Each line is assigned a logistic fragility curve. The failure probability is

$$p_{i,t}^{\text{fail}} = \frac{1}{1+\exp[-(v_{i,t}-V_{50,i})/\beta_i]}, \quad (3)$$

where $V_{50,i}$ is the wind speed corresponding to 50% failure probability and $\beta_i$ controls the slope. A line is disconnected when its probability reaches the specified threshold. The resulting failed lines are expanded into physical repair series so that the scheduler repairs the affected feeder sections consistently.

### B. Reliability Driven Energized ALNS

The classical restoration baseline uses an energized graph decoder embedded in ALNS. A chromosome specifies truck starting warehouses, crew starting warehouses, and crew counts assigned to repair series. The decoder repeatedly identifies productive candidate repairs, estimates their restoration benefit, simulates resource travel, and commits a feasible stage batch. The ALNS outer loop perturbs the chromosome using multiple destroy and repair operators and retains improved schedules.

The objective function is

$$J = w_D\,\text{SAIDI} + \epsilon_F\,\text{SAIFI} + w_E\,\text{ENS} + w_T\,T_{\max} \quad (4)$$

where $T_{\max}$ is restoration makespan. In this study, $w_D = 1$, $w_E = 0.1$, $w_T = 0.001$, and $\epsilon_F = 0.001$. The SAIFI weight is intentionally negligible because the storm induced interruption topology is fixed for a given scenario; scheduling primarily changes outage duration and unserved energy.

The system average interruption duration index (SAIDI), system average interruption frequency index (SAIFI), and energy not supplied (ENS) are calculated using the distribution reliability definitions in [13].

### C. Energized Candidate Pool and Adaptive Stage Filling

The repair series abstraction reduces line level damage into a smaller set of physically interpretable jobs. Each series retains the damaged mapped lines, maximum productive crew count, repair duration, geographic centroid, and material requirement. The energized graph is updated after each committed repair event. Candidate generation begins from jobs that are productive with respect to the currently energized network and also retains useful downstream combinations that can be prepared in parallel. This distinction is important because a downstream repair can be valuable even when it does not immediately restore customers by itself.

The stage fill mechanism is not implemented as a rigid sequence in which all direct jobs must be exhausted before any indirect work is considered. Instead, direct jobs, downstream jobs, and multi job combinations are admitted into a common pool when they help occupy available crews or unlock later service restoration. If the current pool cannot absorb the productive crew capacity, the search boundary is expanded and additional compatible repair atoms are added. This design preserves the original logistics behavior while allowing the binary optimizer to compare a wider range of coordinated stage batches.

### D. QUBO Assisted Stage Batch Selection

The QUBO layer does not replace ALNS or the deterministic decoder. It replaces only the local stage batch selection decision. At a restoration stage, the energized frontier module generates a set of feasible atoms $\mathcal{B}$. An atom may be an individual repair job or a multi job combination. For each atom $b$, the graph evaluator estimates the newly reachable load $\Delta P_b$, newly restored customers $\Delta N_b$, and completion time $\tau_b$. The local utility is

$$U_b = w_E^q \frac{\Delta P_b}{\tau_b} + w_D^q \frac{\Delta N_b}{\tau_b} - w_T^q \tau_b \quad (5)$$

Let $z_b \in \{0,1\}$ denote whether atom $b$ is selected. The local sampler minimizes

$$H(\mathbf{z}) = \sum_{b\in\mathcal{B}} (-\bar{U}_b - \lambda_c\kappa_b + \lambda_a)\, z_b + \sum_{b<c} (\lambda_o O_{bc} - \lambda_s S_{bc})\, z_b z_c \quad (6)$$

where $\bar{U}_b$ is normalized utility, $\kappa_b$ is useful crew capacity contribution, $\lambda_a$ is a mild atom penalty, $O_{bc}$ identifies overlapping repair atoms, and $S_{bc}$ rewards positive graph synergy. The overlap penalty prevents duplicate repair assignments, while the synergy term favors combinations that reconnect downstream loads more effectively together than separately.

Accordingly, the QUBO layer does not automatically favor longer repair chains. It selects individual repairs or multi job combinations according to their restoration contribution, crew capacity value, graph synergy, and physical compatibility. This distinction is important because an individual upstream repair may restore substantial load immediately, while a coordinated downstream batch may be more valuable when several jobs must be completed together to unlock service restoration.

The local QUBO contains a soft atom count bound to control CPU sampling cost. All feasible single job atoms are preserved; if the number of productive single jobs exceeds the nominal atom limit, the effective bound is expanded rather than deleting feasible work. Multi job atoms are then ranked by graph based utility and admitted until the remaining budget is filled. This keeps the QUBO compact while retaining the most relevant parallel restoration alternatives.

A sampled binary solution is not accepted directly. The candidate union is passed through the original deterministic logistics decoder. The productive crew target is calculated from the complete expanded feasible pool:

$$C_t^{\text{target}} = \min\left(C_t^{\text{avail}}, \sum_{j\in\mathcal{J}_t^{\text{pool}}} C_j^{\max}\right) \quad (7)$$

A sampled stage is accepted only when its assigned crews meet this target. This rule prevents productive crews from remaining idle while feasible direct or indirect repair work is available. When the current pool cannot absorb available crews, the candidate boundary is expanded adaptively and the QUBO is rebuilt. When no sampled union passes the deterministic crew truck filter, the algorithm applies the classical stage fill rule as a safe fallback.

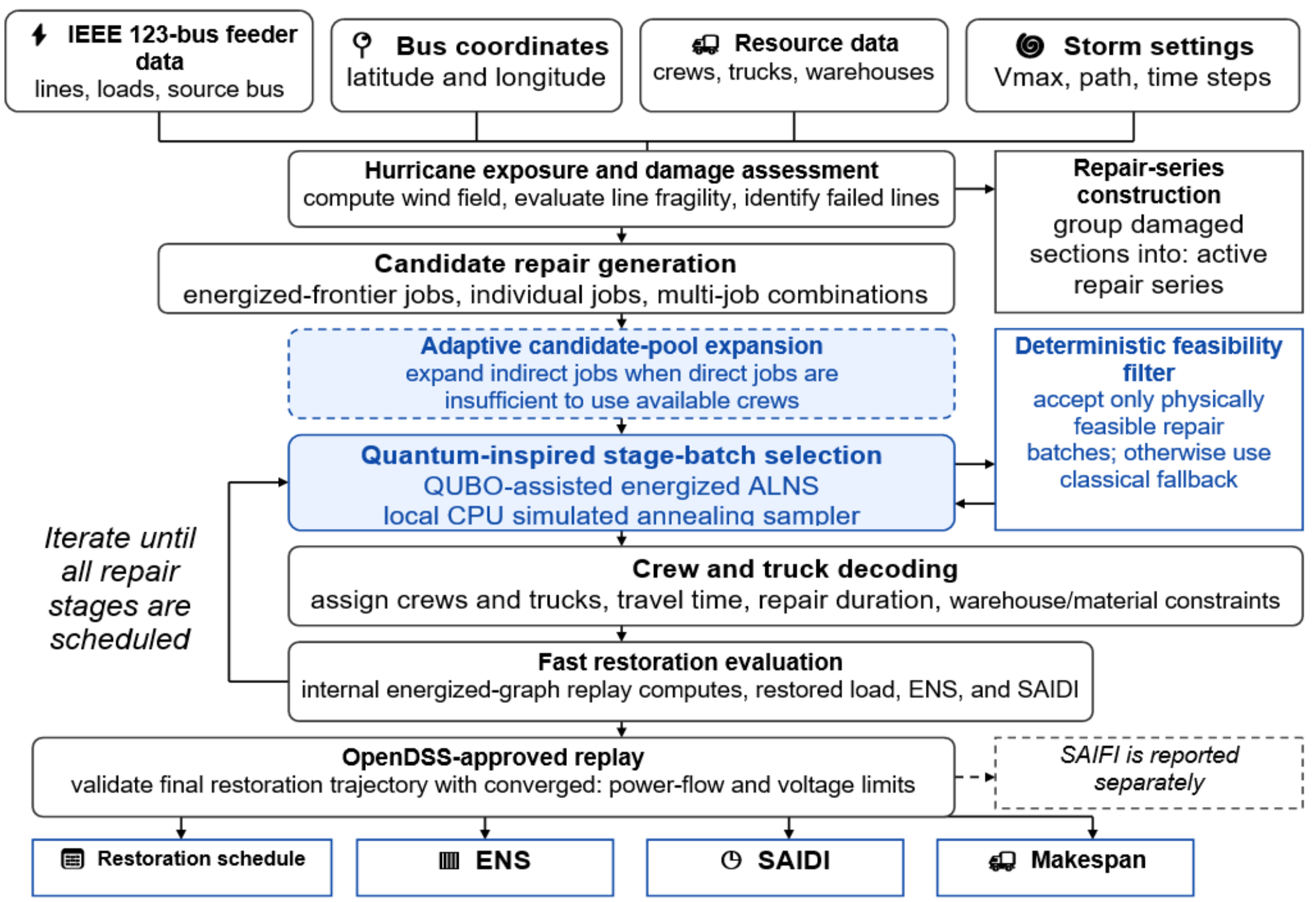

Fig. 1. Overview of the proposed quantum inspired QUBO assisted hurricane restoration framework.

**Algorithm 1: QUBO-Assisted Energized Restoration Scheduling**

**Input:** Damaged repair series, feeder graph, warehouse data, and crew-truck states

**1. Update energized graph** after completed repairs

**2. Generate productive** individual jobs and compatible multi-job atoms

**3. Expand candidate boundary** when useful crew capacity remains available

**4. Build and sample local QUBO** Pass low-energy batches to deterministic decoder

**5. Reject overlapping, infeasible, or under-filled batches** Commit best feasible batch

All repairs scheduled? No → Repeat; Yes →

**6. Replay final restoration** trajectory in OpenDSS and approve schedule

**Output:** OpenDSS-approved restoration schedule

*Fig. 2. Stage level QUBO assisted energized restoration scheduling procedure. Low energy sampled batches are passed to the deterministic crew truck decoder, and only physically feasible batches are committed.*

The stage level procedure is summarized as follows: (1) update the energized graph; (2) generate individual and multi job atoms; (3) expand the candidate boundary when productive crew capacity remains unused; (4) construct and sample the QUBO; (5) pass low energy samples through the deterministic logistics decoder; (6) accept the highest utility feasible batch satisfying (7); and (7) commit the selected repairs and repeat until all repair series are complete.

### *E. Crew Truck Logistics and OpenDSS Validated Replay*

The truck model remains unchanged from the classical decoder. Each truck starts at a selected safe warehouse, travels to an assigned repair location, and may perform additional material round trips when the required poles exceed truck capacity. A repair cannot begin until its assigned crews and truck have arrived:

$$T_j^{\text{start}} = \max\left(T_j^{\text{crew}}, T_j^{\text{truck}}\right) \tag{8}$$

$$T_j^{\text{finish}} = T_j^{\text{start}} + T_j^{\text{repair}} + T_j^{\text{extra trips}} \tag{9}$$

After the optimized schedule is generated, each repaired feeder state is replayed in OpenDSS [14]. A load is counted as restored only after the corresponding feeder solution converges and its voltage satisfies the minimum service threshold. In the reported runs, the admissible range is 0.90–1.10 p.u. Let $a_{l,t}$ denote the service acceptance indicator for load $l$ after restoration event $t$. The OpenDSS validated replay applies

$$a_{\ell,t} = \chi_{\ell,t}\, q_t\, h_{\ell,t} \tag{10}$$

where $\chi_{l,t}$ is the source reachability indicator, $q_t$ is one when the OpenDSS solution converges, and $h_{l,t}$ is one when $0.90 \le V_{l,t} \le 1.10$. Otherwise, the corresponding indicator is zero.

The accepted customer count and unserved demand are recalculated after every committed completion event. This keeps the fast internal graph evaluator available inside ALNS while reserving the full unbalanced OpenDSS solution for schedule approval.

### *F. Local CPU Sampling and ALNS Integration*

The local sampler is used as an inner decision aid rather than as a replacement for the complete restoration search. For each candidate chromosome generated by ALNS, the deterministic decoder updates resource states and constructs the current feasible repair pool. The QUBO sampler then returns low energy atom selections for the present stage. The deterministic decoder re simulates those selections using the original crew and truck states, rejects conflicting or under filled batches, commits the highest utility physically feasible option, and advances to the next restoration stage. The completed schedule is evaluated by the internal reliability model and returned to the ALNS acceptance logic.

This decomposition separates three roles. The outer ALNS loop explores warehouse and crew allocation decisions across complete schedules. The local QUBO layer compares combinatorial repair batches at the current restoration state. The deterministic logistics and OpenDSS layers retain operational feasibility. The present implementation uses CPU simulated annealing so that the formulation can be debugged reproducibly before evaluating cloud based hybrid quantum classical solvers. The QUBO size is controlled by the admitted atom pool rather than by the total feeder size. In the reported implementation, the nominal atom bound is 28. Feasible single job atoms are not discarded when their count exceeds that value; instead, the effective bound is expanded. This safeguard is important because removing productive single repairs solely to meet an artificial software limit could bias restoration results. The same formulation can later be submitted to alternative BQM compatible or hybrid constrained solvers without changing the surrounding restoration architecture.

### *G. QUBO to Ising Compatibility*

The stage selection model is expressed as a binary quadratic model (BQM) and can be mapped directly to an Ising energy model by applying $x_b=(1+s_b)/2$, where $s_b \in \{-1,+1\}$. Under this transformation, atom rewards become local fields, while pairwise overlap and synergy terms become spin couplings. The current CPU sampler evaluates the QUBO form locally; the equivalent Ising representation provides a direct path for future annealing compatible and hybrid solver studies.

**QUBO-to-Ising Mapping for Repair-Atom Selection**

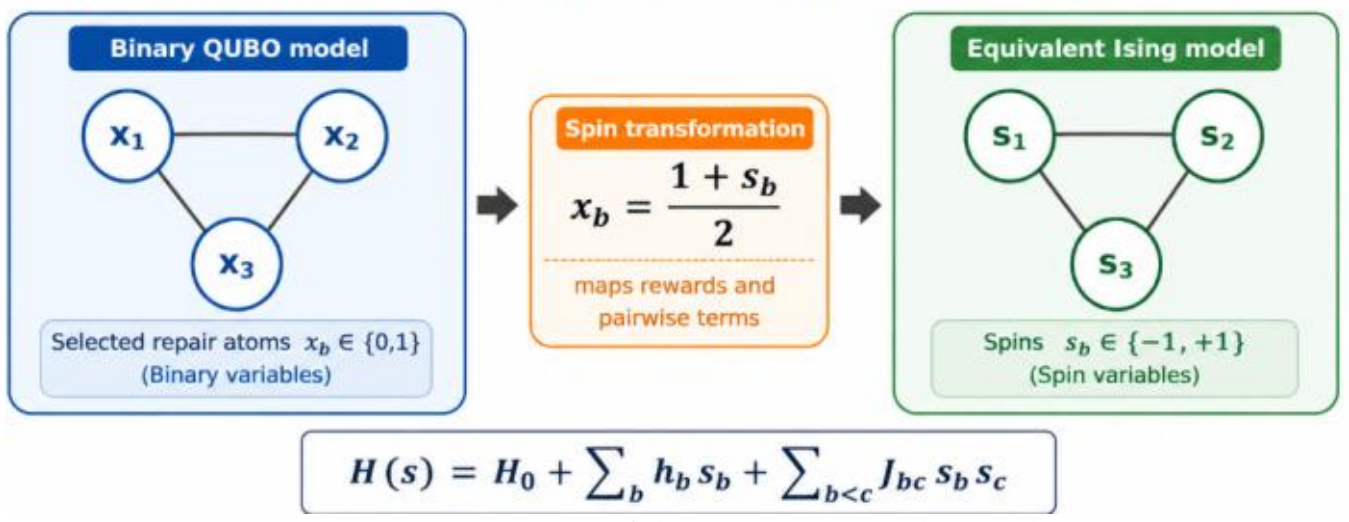

*Fig. 3. QUBO to Ising representation of the repair atom selection layer. Binary repair decisions are mapped to Ising spins, while pairwise overlap and synergy terms become spin couplings.*

## III. Simulation Setup

The no PV IEEE 123 node test feeder is used as the test network. The OpenDSS model contains 126 parsed lines, 91 loads, 698 customers, and 3.49 MW of nominal demand. Coordinates are read from the feeder coordinate file and retained for hurricane exposure, line midpoint calculations, and warehouse placement. The restoration system includes one main warehouse, ten temporary warehouses, 100 repair crews, and ten trucks. Truck and crew speeds are 3 km/h and 10 km/h, respectively. Each method is run for 100 ALNS iterations. The QUBO assisted method uses 300 simulated annealing reads and 1200 sweeps per call. Table I summarizes the cases.

TABLE I: IEEE 123 NODE SIMULATION SETTINGS AND DAMAGE LEVELS

| Item | 80 m/s | 90 m/s | 100 m/s |
|---|---|---|---|
| Disconnected lines | 4 | 10 | 23 |
| Active repair series | 4 | 10 | 22 |
| Expanded mapped lines | 10 | 19 | 40 |
| Crews / trucks | 100 / 10 | 100 / 10 | 100 / 10 |
| ALNS iterations | 100 | 100 | 100 |

The 80 m/s case is a limited damage case. The 90 m/s case introduces a moderate combinatorial scheduling problem. The 100 m/s case is an extreme stress test in which all customers are initially interrupted and 22 repair series must be coordinated. Additional stochastic evaluations are performed for the more severe cases to examine consistency under different search trajectories.

Fig. 4 shows the feeder level line risk map for the 100 m/s case. The spatial distribution of fragility values produces both upstream and downstream repair requirements. In this stress test, damage close to the source creates an initial bottleneck, while distributed downstream failures create a larger parallel scheduling problem after the bottleneck is removed.

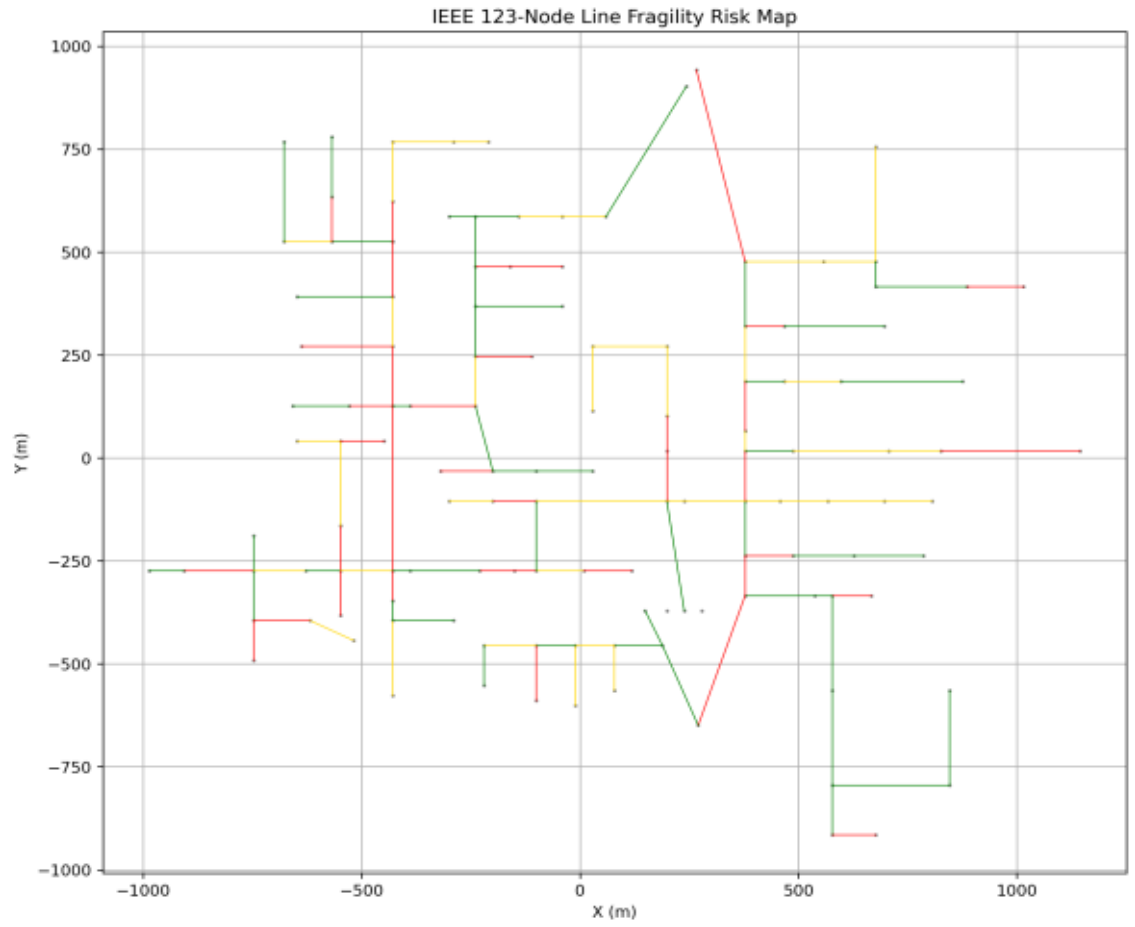

Fig. 4. IEEE 123 node line fragility risk map for the 100 m/s stress test scenario. Colors indicate line level failure risk categories used during damage generation.

TABLE II: LOCAL QUBO SETTINGS USED IN THE CPU EXPERIMENTS

| Parameter | Value |
|---|---|
| Sampler | CPU simulated annealing |
| Reads / call | 300 |
| Sweeps / read | 1200 |
| Nominal atom bound | 28 |
| Validated samples / call | 100 |
| Overlap penalty | 12.0 |
| Capacity reward | 0.35 |
| Synergy weight | 0.35 |

For reproducibility, a line is disconnected when its logistic failure probability reaches 0.70; pole replacement material trips are counted when the corresponding probability reaches 0.80.

Line specific V50 and $\beta$ values are read from the persistent line asset table, which is held fixed across scheduler comparisons. The local QUBO coefficients are $\lambda_o=12.0$, $\lambda_c=0.35$, $\lambda_a=0.03$, and $\lambda_s=0.35$. Atom utility uses ENS, SAIDI, and completion time weights of 1.0, 0.15, and 0.03, respectively. Runs use Python 3.11 with dimod>=0.12 and dwave samplers>=1.2 through the Windows OpenDSS COM interface.

## IV. Results and Discussion

### A. Reliability and Restoration Performance

Table III compares classical energized ALNS with the local QUBO assisted method. At 80 m/s, the classical method performs better in SAIDI and ENS because only four repair series are active and the small search space is already handled effectively by the classical selector. The QUBO assisted method slightly shortens makespan but does not improve the reliability objective. At 90 m/s, the QUBO assisted method improves mean SAIDI and ENS by 1.47% and reduces average makespan by 1.12%. At 100 m/s, the larger scheduling space makes coordinated batch selection more valuable: mean SAIDI and ENS improve by 2.24%, and makespan decreases by 50.71%.

SAIFI remains unchanged between the two schedulers within each storm scenario because the initial hurricane induced topology determines the number of interrupted customers, whereas the repair scheduler primarily changes outage duration and unserved energy.

Fig. 5 illustrates a representative OpenDSS validated restoration trajectory under the 100 m/s stress test scenario. The classical scheduler restores some customers slightly earlier after the source side bottleneck is cleared, but it leaves a long final restoration tail. The QUBO assisted schedule coordinates the remaining repair batches more effectively and completes restoration in approximately 3.50 h instead of 7.10 h.

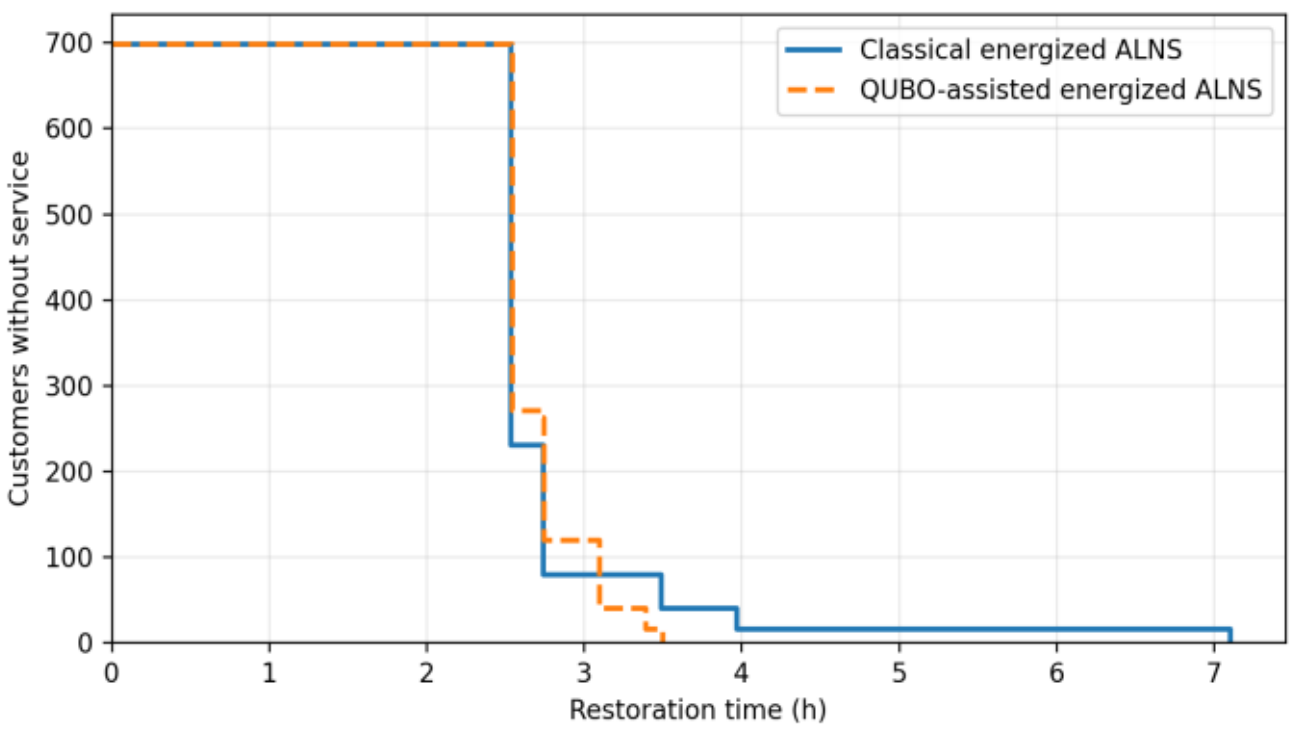

Fig. 5. OpenDSS validated customer outage trajectories for classical and local QUBO assisted energized ALNS under a representative 100 m/s stress test scenario.

TABLE III: CLASSICAL AND QUBO ASSISTED RESTORATION RESULTS

| Wind | Method | SAIDI (h/cust.) | ENS (MWh) | Makespan (h) |
|---|---|---|---|---|
| 80 | Classical | **0.1885** | **0.6578** | 3.3914 |
| 80 | QUBO | 0.2023 | 0.7060 | **3.3677** |
| 90 | Classical | 0.2247 ± 0.0264 | 0.7840 ± 0.0923 | 3.4306 ± 0.0680 |
| 90 | QUBO | **0.2215 ± 0.0278** | **0.7729 ± 0.0972** | **3.3914 ± 0.0000** |
| 100 | Classical | 2.7876 ± 0.0021 | 9.7287 ± 0.0074 | 7.0996 ± 0.0000 |
| 100 | QUBO | **2.7251 ± 0.0191** | **9.5107 ± 0.0668** | **3.4996 ± 0.0000** |

*Note: Values for the 90 and 100 m/s scenarios are reported as mean ± standard deviation across three repeated stochastic evaluations; the 80 m/s case is a single limited damage run.*

### B. Feasibility, Crew Utilization, and Runtime

The severe damage scenario illustrates the adaptive behavior of the scheduler. Initially, only one source critical repair is productive, so the useful target is limited to ten crews. Once this bottleneck is removed, the energized boundary expands and the scheduler assigns all 100 available crews across parallel repair activities. In the final stage, the remaining feasible jobs provide 80 productive crew positions, all of which are assigned. This progression distinguishes the proposed policy from a rigid direct job first strategy.

The deterministic logistics layer also provides a safety mechanism. Fallback warnings occurred only during a limited number of intermediate ALNS candidate evaluations. The final accepted schedules remained physically feasible. OpenDSS replay converged for every reported restoration state, and the minimum service voltage remained above 0.90 p.u.

The quality improvement comes with a runtime tradeoff. The local QUBO method requires approximately 201 s on average at 90 m/s and 361 s at 100 m/s, compared with approximately 16 s for classical ALNS. The current implementation is therefore intended for offline post event planning rather than fast protective control. Its value is most visible under severe damage, where several minutes of computation can be justified by a substantial reduction in restoration tail duration.

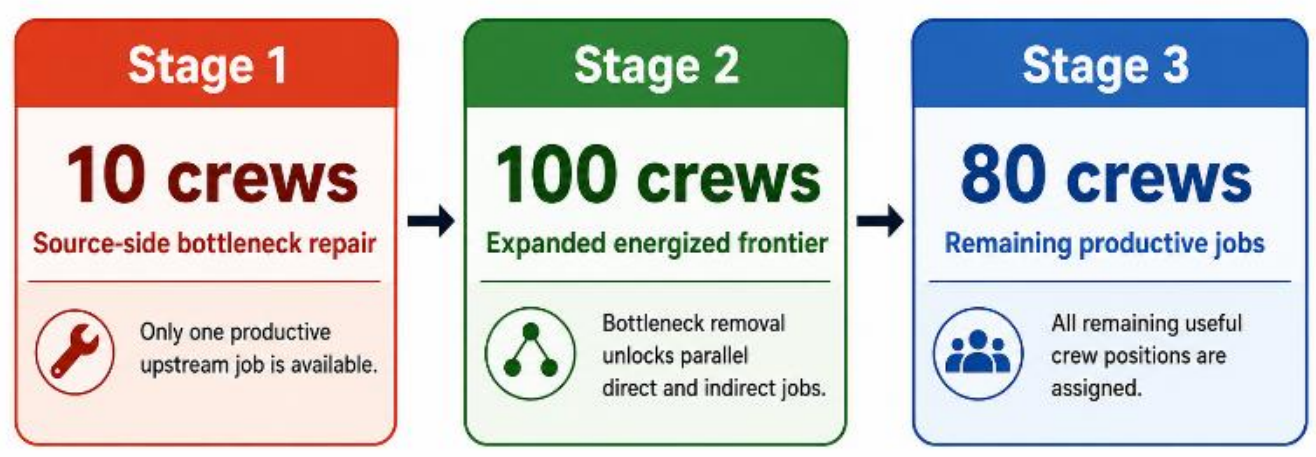

*Fig. 6. Productive crew utilization in the 100 m/s stress test schedule. Repairing the source side bottleneck expands the energized frontier and enables assignment of all useful crew positions in the subsequent stages.*

### C. Damage Dependent Value of QUBO Assistance

The three wind levels reveal a transition in scheduling difficulty. At 80 m/s, only four repair series are active, and the classical selector already performs effectively. At 90 m/s, the larger candidate pool produces modest reliability and makespan improvements. The strongest benefit appears at 100 m/s, where 22 repair series and 40 expanded mapped lines must be coordinated. In that case, the QUBO assisted layer removes the long restoration tail observed in the classical schedule while satisfying the productive crew target shown in Fig. 6. These results indicate that QUBO assistance becomes more valuable as the number of compatible repair batches increases.

### D. Positioning Relative to Existing Quantum Restoration Studies

Recent quantum inspired power system restoration research has primarily focused on topology reconfiguration and microgrid formation. In contrast, the present work addresses a different operational layer: the coordination of physical repair activities after hurricane induced damage. The QUBO model is used only for stage level repair batch selection, while the established logistics decoder retains authority over crew assignment, truck travel, material delivery, and repair timing. This separation allows the binary optimization layer to explore coordinated repair combinations without replacing field operational constraints or the final electrical validation process. The resulting formulation provides a practical intermediate step between classical restoration heuristics and future cloud based hybrid quantum classical implementations. Related distribution system studies by the authors have examined multi objective feeder reconfiguration with DGs and transformer tap changing, as well as radial to ring operation under DG integration [15], [16]. These studies motivate retaining network aware electrical verification while the present work targets the distinct physical repair batch scheduling layer.

Adjacent work has coupled combinatorial search with engineering feasibility checks for protection coordination [17]. Complementary quantum computing studies have embedded QUBO binary blocks within hybrid architectures; a distributed variational quantum eigensolver has been integrated with three block ADMM for unit commitment [18]. Although these problems differ from outage restoration, they support future hybrid solver extensions of the present stage level QUBO.

TABLE IV: POSITIONING RELATIVE TO EXISTING QUANTUM RESTORATION WORK

| Feature | Prior study [7] | Proposed framework |
|---|---|---|
| Primary layer | Microgrid formation and topology reconfiguration | Hurricane repair batch scheduling |
| Model form | Hybrid CQM / quantum annealing compatible model | Stage level QUBO / BQM |
| Evaluated solver | D Wave hybrid quantum classical solver | Local CPU simulated annealing |
| Field logistics | Not the primary focus | Crew, truck, warehouse, and material trip constraints |
| Electrical verification | Distribution operating constraints | OpenDSS validated replay |
| Primary purpose | Restore prioritized loads through microgrid formation | Reduce outage duration and restoration tail after physical damage |

## V. Conclusion and Future Work

This paper introduced a quantum inspired QUBO assisted ALNS framework for post hurricane distribution system restoration. The QUBO layer ranks individual repairs and multi job combinations while a deterministic logistics decoder preserves crew, truck, material trip, and repair time constraints. An adaptive candidate boundary and a hard full useful crew rule prevent unnecessary resource idling. OpenDSS validated replay confirms the electrical feasibility of the resulting schedules.

The results show that QUBO assistance is damage dependent. It is not beneficial for every limited damage case, but it improves mean ENS and SAIDI as the repair space becomes more combinatorial. Under the 100 m/s stress test, the method reduces mean makespan by 50.71% and mean ENS by 2.24% relative to classical energized ALNS. Unlike topology only restoration formulations, the proposed architecture isolates the combinatorial repair batch decision while preserving the operational logistics and OpenDSS validation layers required for field feasible implementation. Future work will evaluate the same QUBO formulation using cloud based hybrid quantum classical solvers, extend the network model to distributed energy resources and islanding, and study larger feeders and additional hurricane realizations.